# Skyrmions in magnetic tunnel junctions

Xueying Zhang[1,2], Wenlong Cai[1], Xichao Zhang[3], Zilu Wang[1], Zhi Li[1,2], Yu Zhang[1], Kaihua Cao[1], Na Lei[1,2], Wang Kang[1], Yue Zhang[1], Haiming Yu[1], Yan Zhou[3], Weisheng Zhao[1,2,]*

[1]Fert Beijing Institute, BDBC, School of Electronic and Information Engineering, Beihang University, Beijing, China

[2]Beihang-Goertek Joint Microelectronics Institute, Qingdao Research Institute, Beihang University, Qingdao, China

[3]School of Science and Engineering, The Chinese University of Hong Kong, Shenzhen 518172, China

*E-mail: weisheng.zhao@buaa.edu.cn

# ABSTRACT

In this work, we demonstrate that skyrmions can be nucleated in the free layer of a magnetic tunnel junction (MTJ) with Dzyaloshinskii-Moriya interactions (DMI) by a spin-polarized current with the assistance of stray fields from the pinned layer. The size, stability and number of created skyrmions can be tuned by either the DMI strength or the stray field distribution. The interaction between the stray field and the DMI effective field is discussed. A device with multi-level tunneling magnetoresistance is proposed, which could pave the ways for skyrmion-MTJ-based multi-bit storage and artificial neural network computation. Our results may facilitate the efficient nucleation and electrical detection of skyrmions.

KEYWORDS: *Skyrmions, magnetic tunnel junction, multi-layers structure, Dzyaloshinskii-Moriya interactions, stray field*

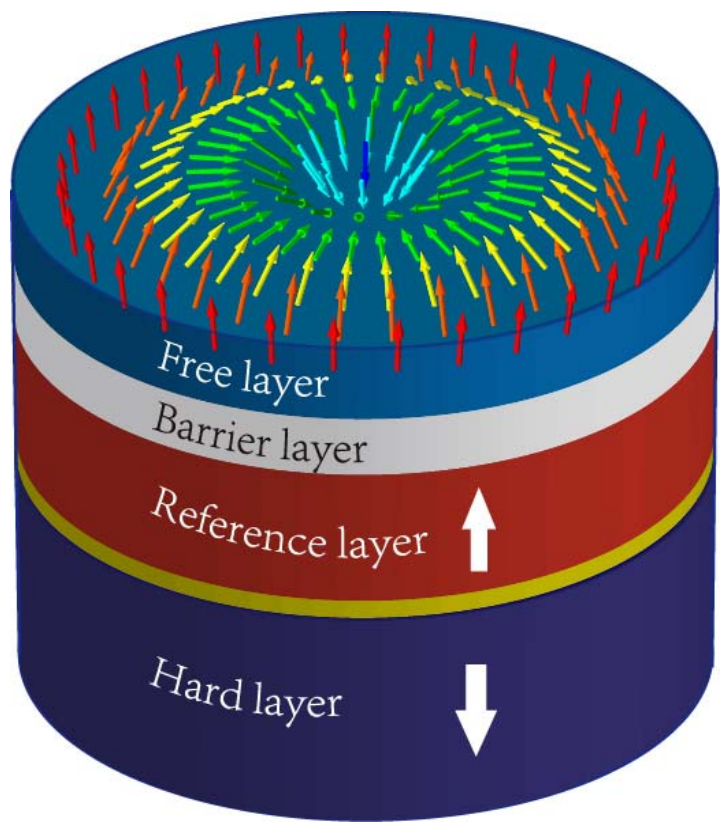

# 1. INTRODUCTIONS

Numerous novel devices for data storage and logic computation have been developed during the past thirty years owing to the emergence of spintronics, among which, the magnetic tunnel junction (MTJ) is one of the most successful examples [1,2]. The prototypical structure of an MTJ is composed of a pinned layer (PL)/barrier layer (BL)/free layer (FL) sandwiched structure. The magnetization in the PL is fixed through an antiferromagnetic structure, while the FL can be switched by the magnetic field, spin transfer torque (STT), or spin-orbit torque (SOT). MTJs can be used to store binary data due to the two opposite magnetic states of the FL. It is also proposed as the synapse for artificial neural network (ANN) computing [3]. However, it is feasible only by considering the stochastic nature of the reversal process in the FL [4] or memristive properties introduced by some strong defects inside or around the device [5]. Tunable and stable intermediate states are strongly desired to obtain higher storage density or more efficient and powerful computing ability.

On the other hand, skyrmions have become one of the hottest research topics in condensed matter physics [6–12] since being first observed experimentally in 2009 [13]. This quasi-particle-like chiral spin texture has many outstanding advantages, such as its nanoscale size and superior stability owing to the topologically protected structure [14,15]. It is also proposed for data storage as an information carrier [16–19] and for the ANN computing as a neurotransmitter [20,21]. However, the nucleation, transport, and detection of skyrmions are still difficult issues. Several nucleation methods have been proposed and demonstrated, for examples, by injecting a localized high-current pulse [22], by coupling with a vortex structure in the adjacent magnetic layer [23], and by exciting a vortex by a current in a special geometry [24,25]. These methods may be difficult to apply either due to the complexity of fabrication or due to the high energy consumption. In addition, skyrmions can only be observed by a few complex microscopic instruments, with which the current cannot be easily applied for further characterization of their dynamic properties. The efficient electrical detection of skyrmions at room temperature is still difficult. The detection of skyrmions by tunneling magnetoresistance (TMR) has been theoretically proposed [26]. However, no idea was proposed regarding the creation of skyrmions in MTJ structures.

An essential condition for the formation of stable skyrmions in a magnetic thin film is the presence of certain Dzyaloshinskii-Moriya interaction (DMI), which is an asymmetric exchange interaction stabilizing non-collinear spin textures. Recent studies suggest that a strong DMI can be generated in the oxide layer/magnetic layer/heavy metal structure [27–30], and stable skyrmions have also been observed in these structures [6,31,32]. Interestingly, this structure is identical to that near the FL of an MTJ: the oxide layer is used as the BL and a heavy metal is usually used as a capping layer (CL) above the FL. A strong perpendicular magnetic anisotropy (PMA) is required both in the FL of a PMA-MTJ and in the materials hosting skyrmions.

In this work, we demonstrate that, if certain DMI exists, skyrmions can be nucleated in the FL of an MTJ during the STT-induced reversal process with the assistance of the stray field $B_S$ from the PL. The effects of the strength of the DMI and the distribution of $B_S$ on the size and stability of skyrmions are discussed. This mechanism enables a prototype device with multi-level TMR due to the skyrmionic state in the FL. This device may solve the issues that hinder the application of MTJs and skyrmions in multi-bit storage, ANN computing, and skyrmion-based oscillators [33].

## 2. RESULTS AND DISCUSSIONS
## 2.1 Creation of skyrmions in MTJs

Figure 1 shows the classical structure of an MTJ: hard layer (HL)/spacer/reference layer (RL)/spacer /spin-polarizing layer (SPL)/MgO/FL/CL. The PL is constructed with a synthetic ferrimagnet configuration for strong coercivity. It consists of three parts: SPL with a thickness of

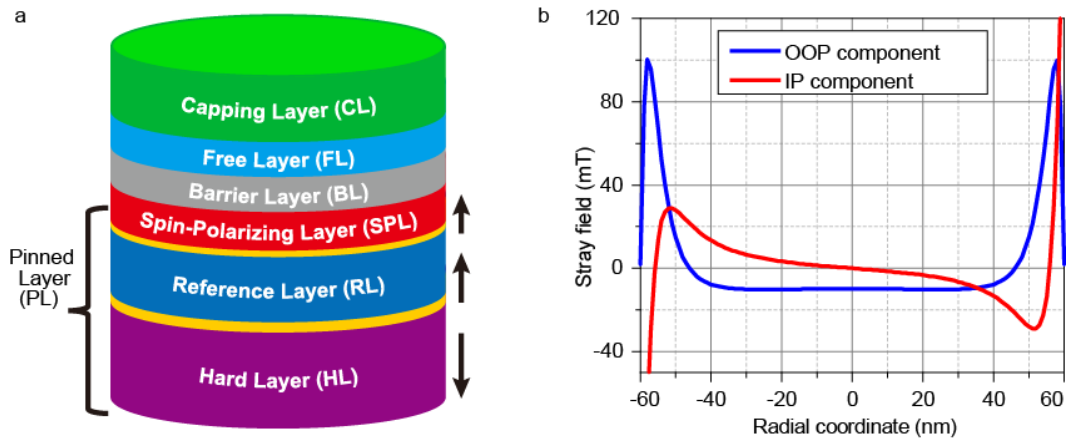

**Figure 1** (a) Multi-layer structure of a classic MTJ. The arrows on the right denote the magnetization of the PL studied in this paper. (b) The space-dependent distribution of the stray field from the PL on the FL in an MTJ with the radius of 60 nm, HL(5 nm)/RL(3 nm)/SPL(1 nm). For $B_{S\perp}$: positive means the +z direction and for $B_{S\parallel}$: positive means radially outward direction.

approximately 1 nm, whose fixed polarization is ensured by a ferromagnetic coupling with the RL through an approximately 0.3-nm-thick metallic. The RL is hardened by an antiferromagnetic coupling with a thicker HL through an approximately 0.8-nm-thick spacer. The HL is usually designed to be very thick to compensate the stray field from the SPL and RL maximally. Note that this stray field will cause a bias of the switching current for the two opposite reversal processes. Through a balancing design, the average net stray field from the PL can be limited to a very low level. However, the stray fields from the SPL, RL, and HL on the FL are not homogeneous. We have calculated the distribution of the stray field from the PL based on a typical MTJ structure, given in Fig. 1a (cf. Supplementary Information (SI) [34] for methods). As shown in Fig. 1b, the perpendicular component of the stray field $B_{S\perp}$ close to the edge of the FL is in the opposite direction to that in the center, with a peak value as large as 100 mT. $B_{S\perp}$ favors a parallel state (P state, i.e., the magnetization of the FL is parallel to that of the SPL) in the edge but an anti-parallel state (AP state) at the center. Moreover, the in-plane component stray field $B_{S\parallel}$ at the center is in the direction opposite to that at the edge of the FL, with a sharp increase at the edge.

Because of the presence of the inhomogeneous distribution of the stray field, the magnetic reversals of the FL in the two opposite directions show different processes: the P-to-AP reversal starts with a domain wall (DW) nucleation from the center, and then expands to the whole FL. On the contrary, the AP-to-P reversal starts with a DW nucleation from the edge, forming a ring-shaped reversed area. Subsequently, the circular DW shrinks, forming a bubble at the center of the FL. Normally, the bubble will collapse and annihilate rapidly due to the surface tension of the DWs and the driving force (e.g. STT). These asymmetric reversal processes have been demonstrated by Gopman et al. through micromagnetic simulations [35]. Through time-resolved studies of the switching of MTJs [36–38], Devolder et al. experimentally found that the AP-to-P reversal takes a longer time than the P-to-AP reversal, and an intermediate state could last for several milliseconds. They speculated that this intermediate state resulted from a bubble formed in the FL.

For a domain bubble with a very small size, the DW surface tension becomes significant. This tension will result in a Laplace pressure, expressed as [39],

$$P = \sigma_{DW}/R \tag{1}$$

where $\sigma_{DW}$ is the DW surface energy and $R$ is the radius of the domain bubble. This pressure leads to the collapse of the bubble and the immediate annihilation of the intermediate state. However, when the DMI exists, the situation changes. On the one hand, the DMI can stabilize chiral spin textures of the unreversed region, namely, forming skyrmions. On the other hand, when the DMI coefficient $D$ further increases, the DW surface energy will be reduced to a very low level, since $\sigma_{DW} = 4\sqrt{AK_{\text{eff}}} - \pi D$ [40] where $A$ is the exchange stiffness, $K_{\text{eff}}$ is the effective anisotropy energy, and $D$ is the DMI constant. In this case, the decreased Laplace pressure can be balanced by the pressure from the stray field, and the skyrmionic bubble is stable or metastable.

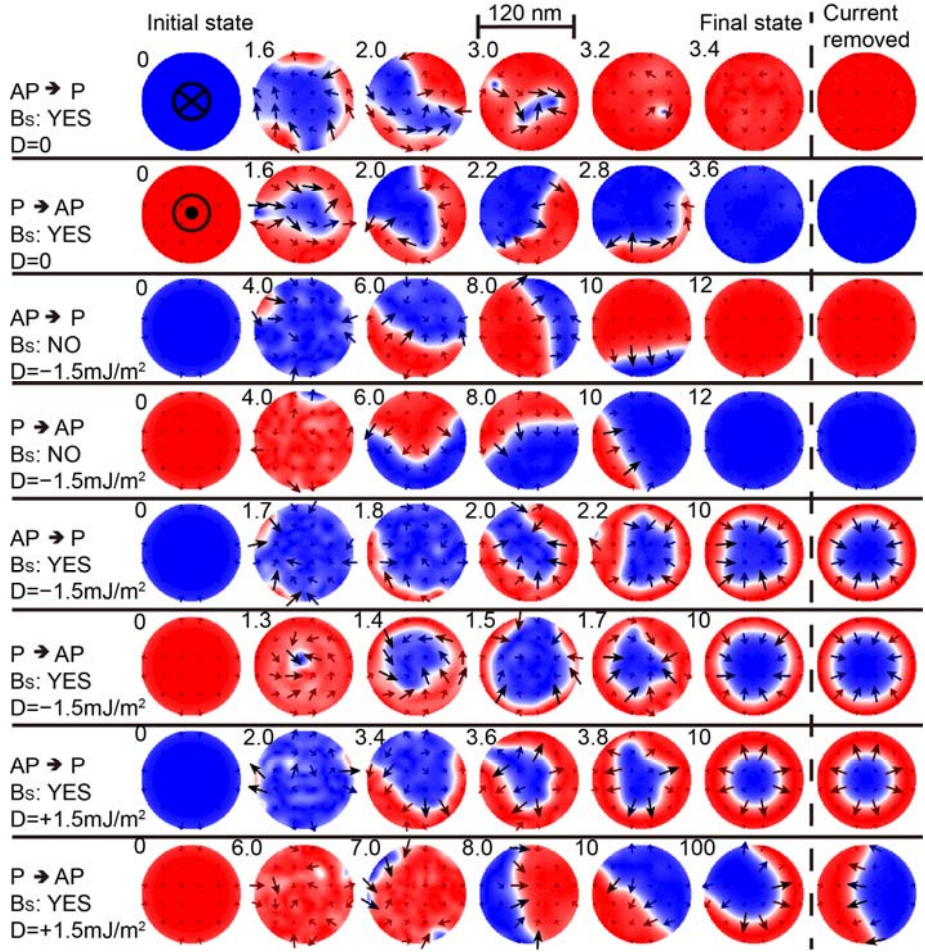

**Figure 2** Formation of skyrmions in the free layer of an MTJ induced by the STT. Applied current density is always $8 \times 10^{10}$ A/m$^2$. Color blue: downward magnetization; red: upward magnetization. The stray field considered here corresponds to that in Fig. 1b. Numbers in the upper left of each image denote the time in nanoseconds. The last column gives the stable state after the applied current is removed.

We simulate the reversal processes of the FL in an MTJ with a micromagnetic simulator Mumax3 [41]. In the simulations [34], the damping-like (i.e., Slonczewski-type) spin-transfer torque (STT) is considered [42,43] and the stray field from the PL is introduced as a space-dependent external field, whose distribution is shown in Fig 1b. Different cases are simulated to check the effects of $B_S$ and DMI on the reversal processes, as shown in Fig. 2.

For the case without $B_S$, both the AP-to-P and P-to-AP reversals start with a DW nucleation from the edge. Subsequently, the DW sweeps the disk, inducing a complete reversal. When the stray field exists as analyzed above, for the AP-to-P switch, the DW first nucleates from the edge and then propagates along the edge with the assistance of $B_{S\perp}$, forming a domain bubble. If no DMI exists, the bubble collapses quickly. However, when the DMI exists, irrespective of the sign of $D$, the bubble can be stabilized, with a certain chirality favored by the effective field $H_{DM}$ of the DMI.

For the P-to-AP reversal, $B_{S\perp}$ favors the DW nucleation first at the center. This analysis is confirmed by the simulation results of the second and sixth rows in Fig. 2, corresponding to zero and negative $D$. For the negative $D$, the reversed domain expands and then a stable skyrmionic bubble forms after several oscillations, whose size is approximately equal to the area covered by the negative $B_{S\perp}$. The bubble remains stable, although the current is applied continuously. This incomplete reversal may be explained by the decreased efficiency of the STT faced with the non-collinear magnetic texture in the FL induced by the DMI [44,45]. The driving force of the STT cannot overcome the Zeeman force of $B_{S\perp}$ at the edge. As expected, for $D = 0$, the reversed domain expands continuously after touching the edge, and finally results in the complete reversal of the FL.

However, for a positive $D$, the P-to-AP reversal first starts from the edge, and no domain bubble is formed. This can be explained by the competition between $B_{S\parallel}$ and $H_{DM}$. Assuming that a small reversed domain with downward magnetization is nucleated at the center, the chirality of the surrounding DW is determined by the sign of the $D$. For a positive $D$, the center magnetization of the DW points outward, opposite to $B_{S\parallel}$ at the center of the FL, as shown in Fig. 1b. In this case, the DW nucleation is hindered by the $B_{S\parallel}$ at the center and hindered by the $B_{S\perp}$ at the edge.

Consequently, the reversal is more difficult. Finally, a complete reversal was not achieved even if the current was applied for more than 100 ns. On the contrary, for a negative $D$, both $B_{S\parallel}$ and $B_{S\perp}$ favor the reversal at the center. This interplay between $B_{S\parallel}$ and $H_{DM}$ is also the reason why the size of the skyrmion obtained in the stable state is different for positive and negative $D$ during the AP-to-P reversal, as can be seen from the last column of Fig. 2. Technically, the sign of $D$ can be tuned by choosing the different heavy-metal materials for the CL in the device design [27,46].

## 2.2 Stability and size of skyrmions in MTJs

To further investigate the role of the DMI in the formation and stabilization of skyrmions, we simulate the magnetic state of the FL by varying $D$ from 0 to −2.5 mJ/m² with different current densities, as shown in Fig. 3. We can see that the threshold current for DW nucleation increases with the strength of the DMI when $D$ is relatively small (less than 1.25 mJ/m² here). This may be explained by the detrimental effect of the DMI on the efficiency of the STT [44,45]. However, the nucleation current decreases when $D$ further increases. This may be explained by the fact that a non-collinear spin texture is closer to the ground state when the DMI is larger than certain threshold value [40]. Based on the parameters used in our simulations, the DW surface energy is lower than a uniformly magnetized state when $D$ is larger than a critical value $D_c = 4\sqrt{AK_{eff}}/\pi = 1.9 \text{ mJ/m}^2$.

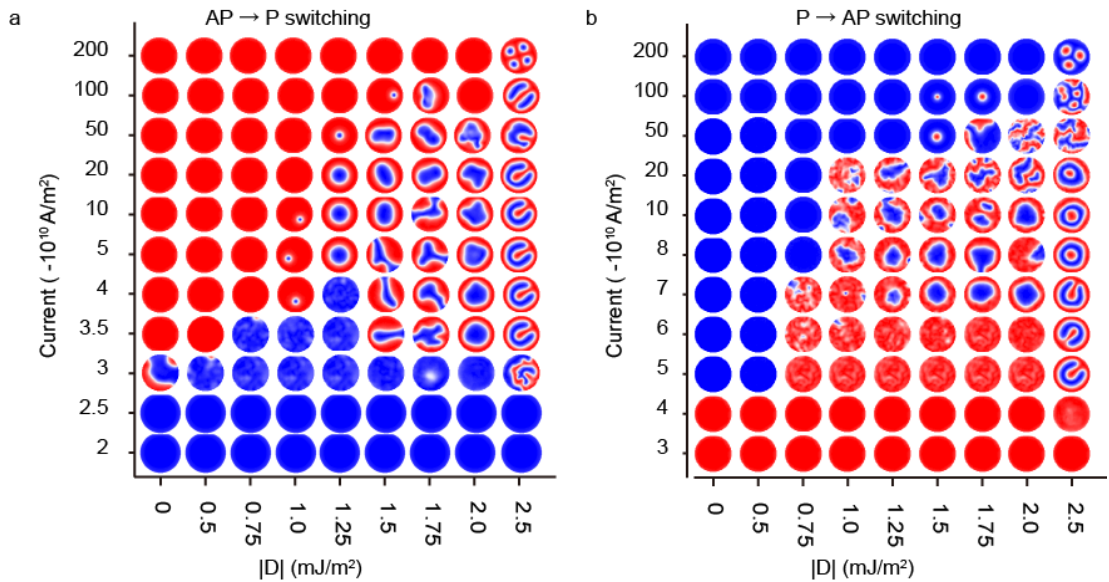

**Figure 3** Magnetic states for different values of $D$ and applied current. Figures are obtained as follows: for each specified $D$, the current was applied continuously, changing from 0 to 2 TA/m² (or to −2 TA/m²) in a stepwise manner. For each step, the current is applied for 10 ns and a snapshot is taken at the end of the duration. The stray field that is considered corresponds to that in Fig. 1b. The sign of $D$ used is negative. The radius of the disk is 60 nm.

For the AP-to-P reversal, a skyrmion can form only when $D \geq 1 \text{ mJ/m}^2$. The size and the stabilization of skyrmions are more sensitive to the value of $D$, but less sensitive to the STT. For $D \approx 1 \text{ mJ/m}^2$, the skyrmion is compact and less stable; for $D \geq 1.5 \text{ mJ/m}^2$, the skyrmion is larger and more stable. In this case, the skyrmion is mainly stabilized by the stray field from the PL and demagnetizing field from the FL itself. The large skyrmion shrinks to a compact skyrmion only when the current increases to a threshold value, and it annihilates when the current further increases. The threshold current for compaction and annihilation increases with $D$. When $D$ further increases, a spin spiral state appears following the DW nucleation. A multi-skyrmion state is obtained when the applied current increases to a very large value.

For the P-to-AP reversal, when $D \geq 1 \text{ mJ/m}^2$, a stable bubble can form after a DW nucleation at the center. However, as the applied current further increases, the magnetization in the FL becomes strongly disordered, which may be caused by the competition between the STT and the strong $B_{S\perp}$ at the edge of the FL. Interestingly, as the applied current further increases, the strongly disordered state is terminated by forming a single skyrmion or multiple skyrmions with a magnetization opposite to that of the bubble formed initially.

In addition to the DMI, the strength and distribution of the stray field, which can be tuned by the structure of the PL, can also be used to manipulate the formation and stabilization of skyrmions in the FL. Based on our further study [34], if a stronger stray field exists, the requirement on the strength of the DMI to obtain skyrmions in the FL can be extended. Although the radius of all the MTJs simulated above is 60 nm, our studies show that skyrmions can be successfully created in MTJs using this mechanism with wide-ranging sizes [34]. However, the condition for the creation of skyrmions in a smaller MTJ is stricter.

**2.3 Detection of skyrmions using TMR**

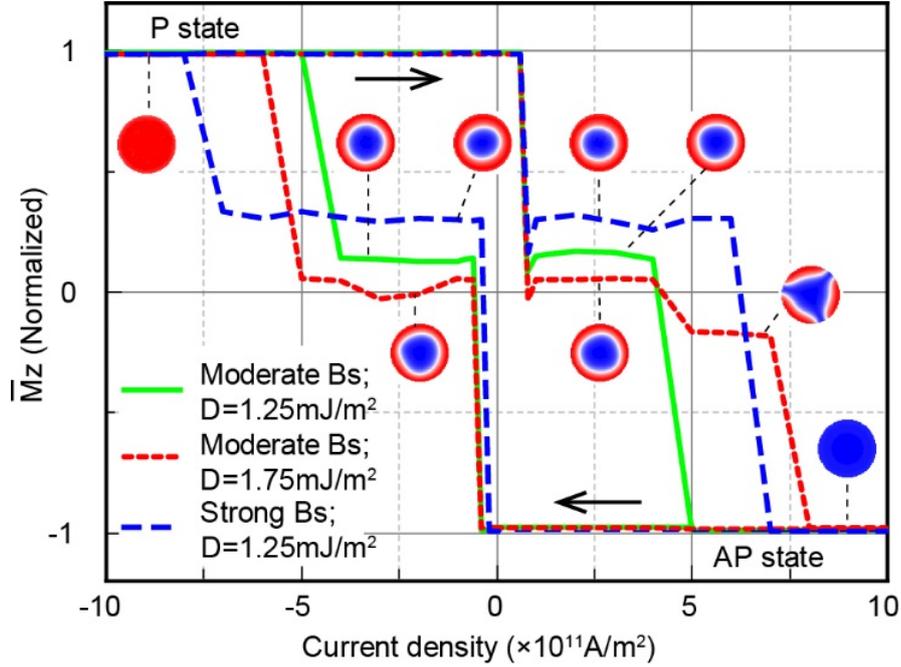

**Figure 4** Magnetization-current hysteresis loops obtained by simulating the switching of $R = 60$ nm MTJs with different $D$ and $B_S$. A moderate $B_S$ corresponds to the configuration in Fig. 1 and a strong $B_S$ is obtained with a thicker PL: HL(5 nm)/RL(10 nm)/SPL(1 nm) (see Supplementary Information). The current flowing from the PL to the FL is defined as positive. The current is changed in a stepwise manner, and each step lasts for 10 ns. Between two steps, the current is stopped for 10 ns and $\bar{M}_z$ is saved at the end of the relaxation. The disks inside give the corresponding magnetic state of the FL.

The resistance of an MTJ is determined by the magnetic state of the FL with respect to that of the RL [42]. If skyrmions exist, an intermediate resistance between the high and low level is obtained. The number of the intermediate states is determined by the number of skyrmions. Once a skyrmion is created in the FL, its properties, such as size and stability, can be electrically detected through TMR.

To check the functionality of this skyrmion-based multi-level device, we simulated a hysteresis current scanning of the FL of MTJ with different $D$ and $B_S$. The average perpendicular magnetization $\bar{M}_z$ was calculated and plotted in Fig. 4, which can be seen as an indicator of the resistive response of the device to an applied current. It can be seen that a stable intermediate resistance appears in both the P-to-AP and AP-to-P reversal processes. The stability and resistance of this intermediate state can be tuned by the strength of the DMI as well as the distribution of the stray field, as discussed above.

The device with a stable skyrmionic state in the FL can be used for multi-bit data storage and for ANN computing. The creation, manipulation, and detection of skyrmions can be achieved in the same MTJ, and no motion is required. Moreover, the controllable and energy-efficient method to nucleate skyrmions described here is helpful for the design of other skyrmion-based spintronic devices.

## 3. CONCLUSIONS

In conclusion, we demonstrated that the stray field from the PL of an MTJ on the FL is spatially inhomogeneous, and in opposite directions at the center and edge. With the assistance of this stray field, skyrmions could be created in both the STT-induced AP-to-P and P-to-AP reversal processes in the FL with certain DMI. The size and stability of skyrmions can be tuned by either the strength of the DMI or the stray field. An MTJ with stable intermediate states can be realized based on the skyrmionic states in the FL, which may be used in multi-bit data storage and ANN computation. The new mechanism for skyrmion creation with low energy consumption will expedite the design and development of skyrmion-based devices.

**Acknowledgement**

W.Z. acknowledges the support by the Chinese Postdoctoral Science Foundation (Grant No. 2015M570024), the National Natural Science Foundation of China (Grants Nos. 61501013, 61471015 and 61571023), Beijing Municipal Commission of Science and Technology (Grant No. D15110300320000), and the International Collaboration Project from the Ministry of Science and Technology of China (Grant No. 2015DFE12880). Y.Z. acknowledges the support by the President's Fund of CUHKSZ, the National Natural Science Foundation of China (Grant No. 11574137), and Shenzhen Fundamental Research Fund (Grant Nos. JCYJ20160331164412545 and JCYJ20170410171958839).

**Supporting information:**

Research methods. Influence of the sign of DMI on the reversal and the formation of skyrmions. Influence of stray field on the formation and stability of skyrmions. Influence of size of MTJs on the formation of skyrmions. Magnetization - current hysteresis loop.

# Supporting information

## I. Detailed information about the research methods

### 1. Calculation of the stray field

The demagnetizing field from the Pinned Layer (PL) was calculated using the concept of magnetization current[1,2]. As shown in Fig. S1, magnetic domains can be infinitely divided into small magnetic elements. Each element is equivalent to a ring current element, with current $I = t_M M_S$, where $t_M$ is the thickness of calculated magnetic layer (Hard Layer (HL), Reference Layer (RL) or Spin-Polarizing Layer (SPL)) and $M_S$ is the corresponding saturated magnetization. For a disk which is uniformly magnetized in the out-of-plane direction, the current of one magnetic element can always be cancelled out by the current of the neighboring ones, except for elements in the edge of the disk. Therefore, the magnetization currents are zero inside the disk. The only non-zero contributions are on the edge of the domain, where there is a line of current parallel to the edge, $I = t_M M_S$. The demagnetizing field from each layer is equal to the Oersted field produced by this current, which had been numerically calculated base on the Biot-Savart law. The total stray field including the in-plane component and the out of plane component was obtained by accumulating the stray field from the HL, RL and SPL.

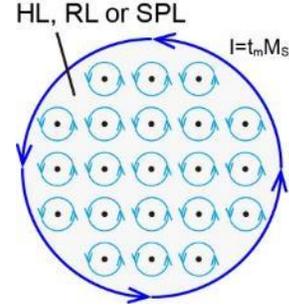

*Figure S1 The effective current along the edge of the a disk which is uniformly magnetized in perpendicular direction.*

In our calculation, the saturation magnetization of the HL, RL and SPL are all considered as $M_S = 1.1 \times 10^6 \, A/m$, the same as the free layer. The FL is in the top of the MTJ by default. The direction of the magnetization of HL is in downward and that of RL and SPL is upward. The thickness of the SPL is considered as 1nm, the same as the Free Layer (FL). Here, we calculated the stray field in three cases, representing the case with weak stray field, moderate stray field and strong stray field:

Case 1: Structure with weak stray field: HL (2.5nm)/RL (1nm)/PL (1nm)

Case 2: Structure with strong stray field: HL (10nm)/RL (5nm)/PL (1nm)

Case 3: Structure with moderate stray field: HL (5nm)/RL (3nm)/PL (1nm)

Results are plotted in Fig. S2. We can see that the stray field can be tuned by changing the thickness of HL and RL. In all our simulations, both the out of plane (OOP) component and in-plane (In-P) component of the stray field are considered simultaneously.

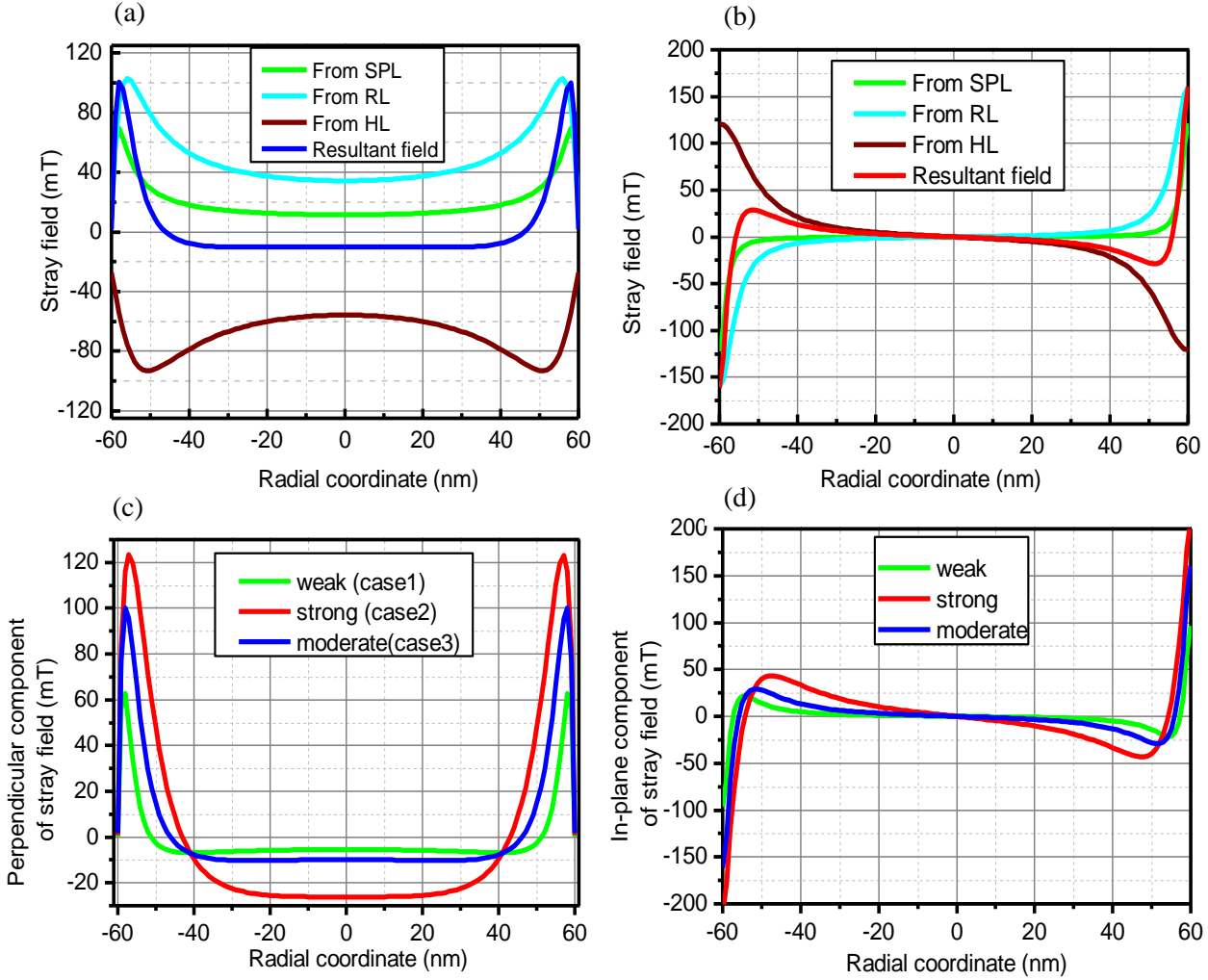

*Figure S2 (a) & (b) Out of plane (in-plane, respectively) component of stray field from the SPL, RL, HL and the total stray field corresponding case 3. (c) & (d), the total out of plane (in-plane) component of the stray field from PL.*

## 2. Micro-magnetic simulations

Micro-magnetic simulations are performed with Mumax3[3]. Only the magnetic evolutions in the free layer were simulated. The thickness of the free layer $t_M$ was always set to be 1nm. The stray field from the PL was simulated by introducing a space-dependent external field, as calculated above. Parameters of the free layer are taken based on the properties of CoFeB thin layer[4]: $M_S = 1.1 \times 10^6 \, A/m$, uniaxial perpendicular anisotropy energy $K_U = 9.0 \times 10^5 \, J/m^3$, exchange stiffness $A_{ex} = 1.0 \times 10^{-11} \, J/m$, damping constant α=0.01, spin polarization P=0.72.

A perpendicular spin-polarized current $j_Z$ was homogenously applied on the whole section of FL. The damping like spin transfer torque (DL-STT) was considered, expressed as[3]:

$$\vec{\tau}_{DL} = \tau_{DL}(\vec{m} \times (\vec{m}_P \times \vec{m})) \qquad (1)$$

$$\tau_{DL} = \frac{\hbar P j_Z}{2M_S e t_M} \qquad (2)$$

where $\vec{m}$ is the magnetization direction in FL, $\vec{m}_P$ is the fixed layer polarization.

As for the field like spin transfer torque (FL-STT), it can be expressed as,

$$\vec{\tau}_{FL} = -\tau_{FL}(\vec{m} \times \vec{m}_P) \tag{3}$$

Now, let us estimate the effective field it produces. If we consider[5–7] $\tau_{FL} \sim 20\% \cdot \tau_{DL}$, and the current applied is $j_z = 1 \times 10^{11} \, A/m^2$. We obtain $\tau_{FL} \sim 4 \, mT$. We can see that this field is relatively weak compared with the stray field. Moreover, the FL-STT shows an even dependence on the bias voltage in symmetric structure (i.e., the sign of the effective field remain the same in the AP to P switch and P to AP switch although the direction of current is changed) [8]. Therefore, the FL-STT was not considered in the simulation. In fact, with the Mumax code, the FL-STT can be eliminated by setting $\varepsilon' = \alpha\varepsilon$, where ε and ε' is the primary and secondary spin-torque parameter.

DMI term is included as an effective field,

$$\vec{B}_{DM} = \frac{2D}{M_S}\left(\frac{\partial m_z}{\partial x}, \frac{\partial m_z}{\partial y}, -\frac{\partial m_x}{\partial x} - \frac{\partial m_y}{\partial y}\right) \tag{4}$$

where *D* is the DMI constant.

At last, we would like to notice that a small thermal fluctuation was introduced in all the simulation to eliminate the DW nucleation induced by the resonant excitation of the spin wave. In the simulation, properties of FL were homogeneous and the shape was symmetric (circle). If no fluctuation were introduced, we found that there would be a spin wave propagating along the radial direction and it oscillates between the edge and the center. This spin wave is strictly symmetrical by the center of the circle and has a synchronized phase. A symmetric DW nucleation will occur after the spin wave superposing in the edge or in the center, as shown in Fig. S3. This case does not exist in the real reversal process since the strict symmetry of the spin wave will be broken by even a small defect in material or by a thermal fluctuation. Therefore, we introduced a very small thermal fluctuation by setting the temperature to 1K. Indeed, the thermal fluctuation is mimicked by a randomly distributed magnetic field in Mumax$^3$.

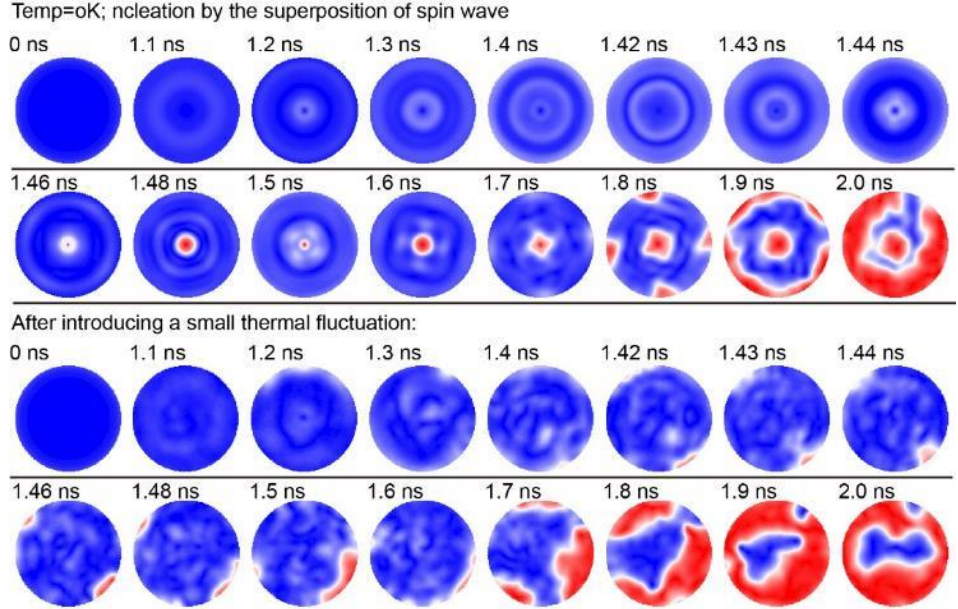

*Figure S3 The AP to P reversal process of a R=75nm MTJ. $D = 1.5 \, mJ/m^2$; $j_z = 8 \times 10^{10} \, A/m^2$. (a) No thermal fluctuation was introduced; (b) a small thermal fluctuation was introduced by T=1K.*

## II. Influence of the sign of DMI on the reversal and the formation of skyrmions

The DMI act as an effective field $H_{DM}$. For a skyrmion or a bubble in the center of the FL, the $H_{DM}$ is always along the radial direction. Since the in-plane stray field is also along the radial direction. It is worth to verify the influence of the sign of DMI on the formation of skyrmions.

In this simulation, a moderate stray field was taken into account, as plotted in Fig. S2 (a) & (b). The applied current density are all $|j_z| = 8 \times 10^{10}\ A/m^2$, the radius of the disk of FL is 60nm. We studied the AP to P and P to AP reversal by changing the sign of D. Results are shown in Fig. S4.

From the simulation results, for negative D, the reversal is faster and a skyrmion can be obtained in both the reversal process. However, for positive D, the reversal takes longer time. In addition, during the P to AP reversal, DWs nucleation start from the edge and no skyrmion is obtained. It can be explained by the relative orientation of the $H_{DM}$ and $B_{S\parallel}$.

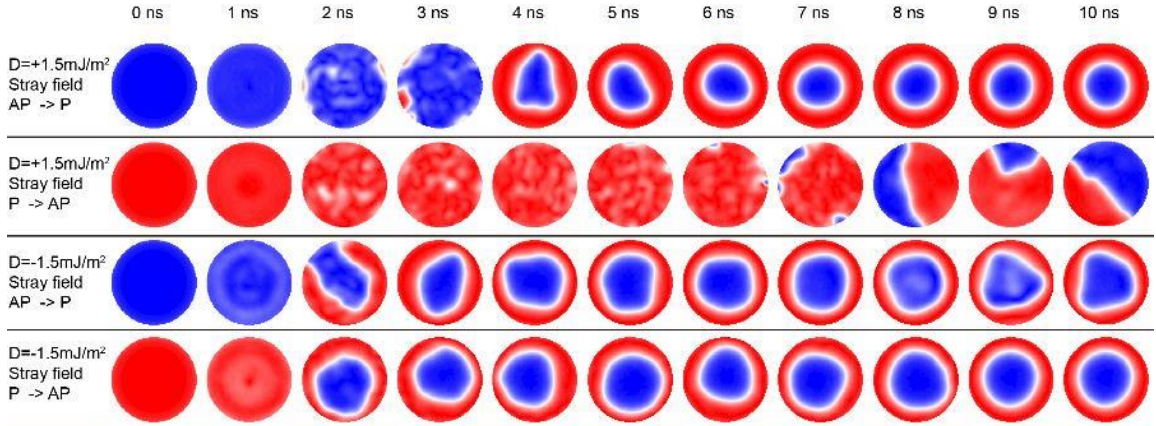

*Figure S4 the magnetic reversal process of an R=60nm FL of a MTJ for different sign of D. $|j_z| = 8 \times 10^{10}\ A/m^2$ for all these cases.*

Fig. S5 (a) and (b) show the stable magnetic state of a R=60nm disk when a skyrmion exists. Neither a stray field nor a spin current was applied. One can see that the magnetization in the center of DW points always to the red part (+z magnetization) if D is positive; vice versa. In the edge, because of the DMI, the magnetization is tilting and the in-plane component points to the red part for positive D. Now, the stray field is taken into account, as shown in Fig. S5 (c) and (d). For the P to AP reversal, the edge was fixed by the $B_{S\perp}$. If a small nucleation occurs in the center, the orientation of the magnetization of the DW around the nucleation point is determined by the sign of D. For positive D, DW magnetization points outwardly, opposite to the direction of $B_{S\parallel}$. Therefore, nucleation in the center is not favored by $B_{S\parallel}$. As a result, the nucleation in this case is relatively difficult. Finally, the DW nucleation starts from the edge and no skyrmion is obtained. On the contrary, for negative D, the nucleation in the center is favored by both $B_{S\perp}$ and $B_{S\parallel}$; the reversal is faster than the case of positive D.

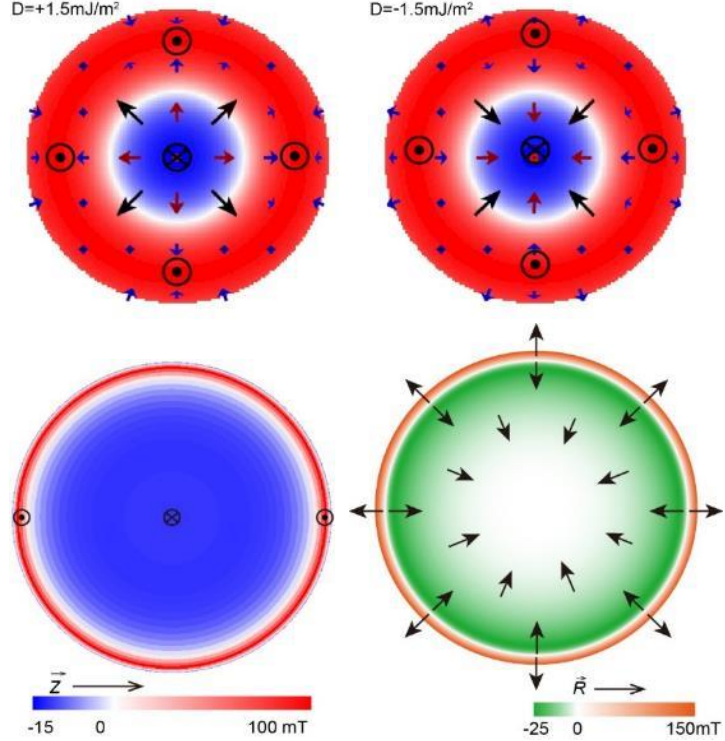

*Figure S5 (a) the stable magnetization of a R=60nm disk with a skyrmion in the center in nature sate (i.e. neither external field nor current applied) for a positive D; (b) negative D. (c) Distribution of $B_{S\perp}$ and (d) $B_{S/\!/}$ in the FL of a MTJ. The stray field displayed here corresponds to the moderate one plotted in Fig. S2.*

## III.   Influence of stray field on the formation and stability of skyrmions

As demonstrated in Fig. S2, the magnitude of the stray field can be tuned by modulating the thickness of RL and HL. A couple of relatively thicker RL and HL gives a stronger stray field (i.e., a larger contrast between the field magnitude in the center and in the edge). Here, we simulated the reversal of the FL for different stray fields. The six phase diagrams were obtained with the same way as that of Fig.3 in the main text except for the difference of stray field.

We can see that skyrmions can be obtained even when the stray field is weak. However, the constraint of D is stricter for a weaker field. For example, when $D=1mJ/m^2$, skyrmion cannot form with a weaker stray field. However, a compact skyrmion was obtained with a moderate $B_S$. A large skyrmion can be obtained with a strong $B_S$. For the AP to P reversal, when the stray field is strong enough, the size of skyrmions is mainly determined by the area covered by the negatively oriented stray field. During the P to AP reversal, it is easier to obtain a reversed bubble in the center of FL with a stronger stray field.

The presence of DMI and STT can result in a disordered magnetic texture[9,10]. For the AP to P reversal, the shape of skyrmions is not regular when D is large. For the P to AP reversal, the disordered texture is more obvious, especially when the applied current increases to 200 GA/m². After this disordered state, a single or more skyrmions can be obtained if D is large enough.

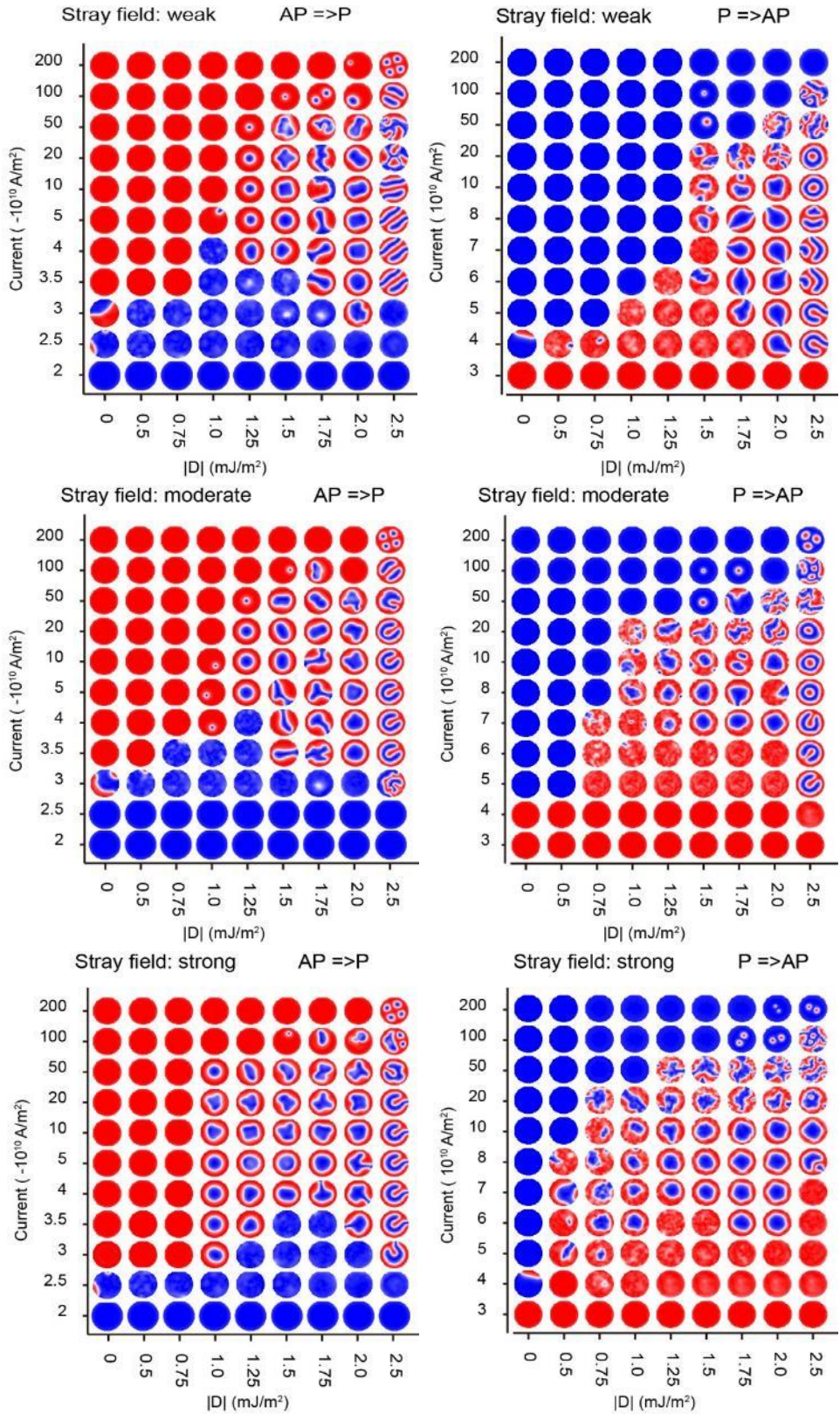

*Figure S6 The magnetic reversal of the FL of R=60nm MTJ with different stray field. The stray field used here corresponds to the calculated results shown in Fig. S2(c) & (d).*

## IV. Influence of size of MTJs on the formation of skyrmions

Most of the above simulations were based on MTJs with a radius of 60 nm. Here, we investigated the formation of skyrmions in the FL of MTJs with different size.

### 1. The magnetic reversal of an MTJ with R=30nm

First, we simulated the reversal of the FL of MTJ with R=30nm. The stray field considered is shown in Fig. S7, which can be obtained by such a configuration: SPL(1nm)/RL(3nm)/HL(6nm). We found that in the AP to P reversal, a skyrmion could be obtained with the same procedure as that in a R=60nm disk: DW nucleation starts from the edge and then the reversed domain surrounds the unreversed part in the center, forming a skyrmion. The skyrmion in the center is very stable. It was not broken until the applied current reached 2.5 TA/m$^2$.

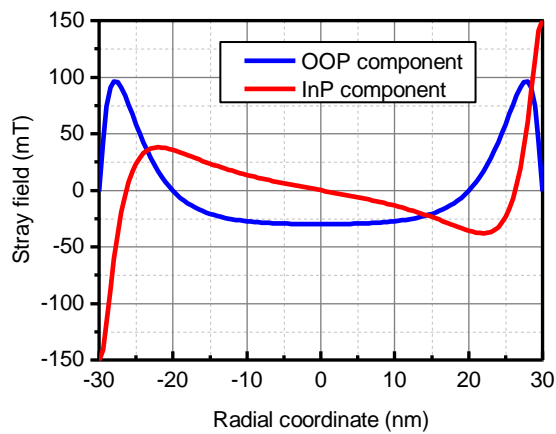

*Figure S7 The distribution of stray field in the FL of a R=30nm MTJ. Structure of PL: SPL(1nm)/RL(3nm)/HL(6nm).*

However, the P to AP reversal becomes complex. The nucleation of reversal in the center becomes difficult. This may be explained by the fact that the magnetization in the center is strongly exchange-coupled with the surrounded zone in the edge, which is fixed by $B_{S\perp}$. In addition, the demagnetization field from the FL itself helps the nucleation. However, this demagnetization field is weaker in a smaller disk than in a larger disk. Finally, the reversal starts by the DW nucleation in the edge. Then, the reversed area propagated to the center. Although a skyrmion can form in the center, however, it was not stable. It will touch the edge soon after a relaxation. A current was continuously applied for more than 100ns and we found that a complete reversal of the FL could not be achieved. The magnetization of FL alternated between an unstable skyrmion in the center and a domain adhering to the edge.

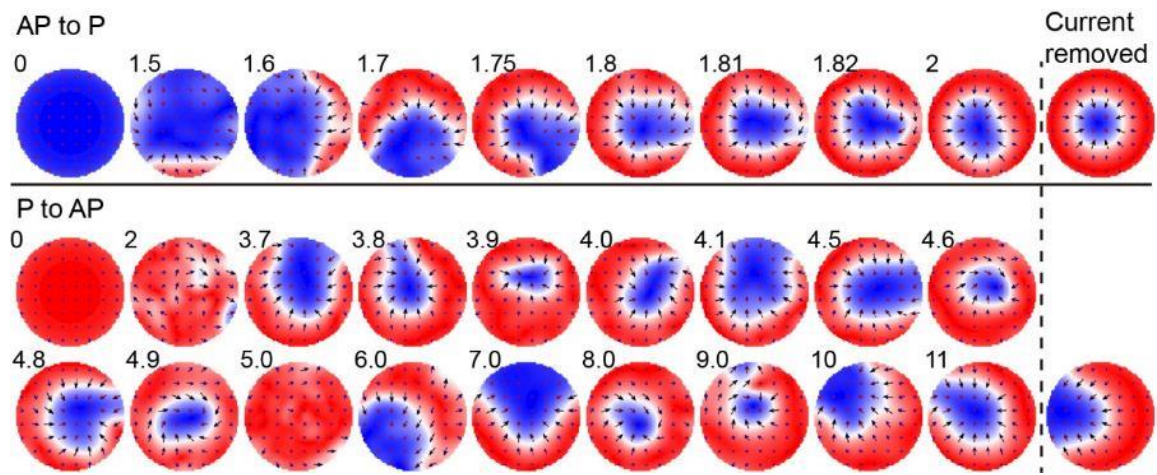

*Figure S8 The magnetic reversal of the FL of a R=30nm MTJ; $D = 1.5 \ mJ/m^2$; for the AP to P reversal: $j_z = -8 \times 10^{10} \ A/m^2$; for P to AP reversal: $j_z = 12 \times 10^{10} \ A/m^2$. Time is given in ns in the upper left of each image.*

Besides, we have tried a stronger stray field, which was obtained by such as configuration: SPL(1nm)/RL(7nm)/HL(14nm). A stronger DMI was also tried. However, it was always difficult to obtain a stable skyrmion in the P to AP reversal.

## 2. The magnetic reversal of an MTJ with R=120nm

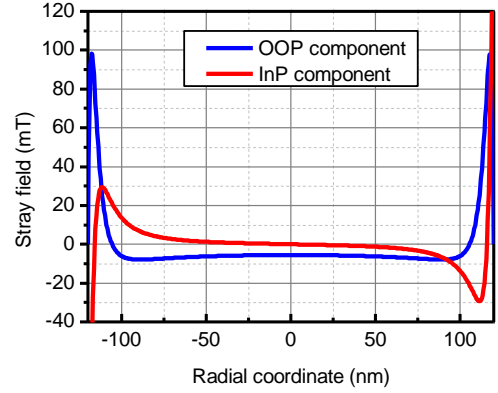

Figure S9 Distribution of the stray field in the FL of a R=120nm MTJ. Structure of PL: SPL(1nm)/RL(3nm)/HL(5nm).

Similar to the reversal process of the R=60nm MTJs, the AP to P reversal of a 120nm MTJ starts from the edge and the P to AP reversal starts from the center. A partially reversed state can be maintained for a long time even when a current is applied continuously. Compared with the R=60 nm MTJ, here, a skyrmionic bubble can be obtained with a relatively weak DMI. For example, with $D = 1.0$ mJ/$m^2$, a stable skyrmionic bubble can form during the reversal of both directions, as shown in Fig. S10. This may be explained by the relatively stronger demagnetizing field from FL itself in a larger MTJ, which helps the stabilization of the bubble in the center.

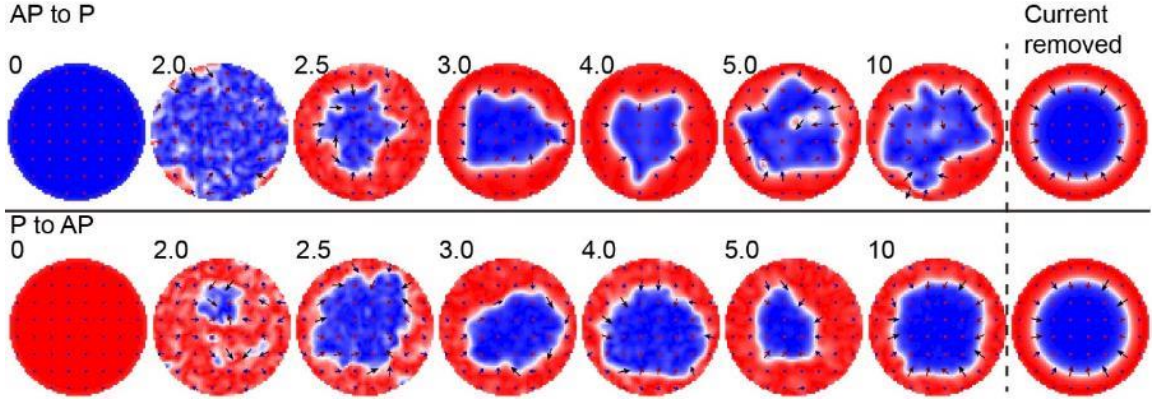

*Figure S10 The magnetic reversal of the FL of a R=120nm MTJ; for both reversals: $|j_z| = 8 \times 10^{10}$ A/m$^2$; D=1.0 J/m$^2$. The space dependent distribution of the stray field used for this simulation is shown in Fig.S9. Time is given in ns in the upper left of each image.*

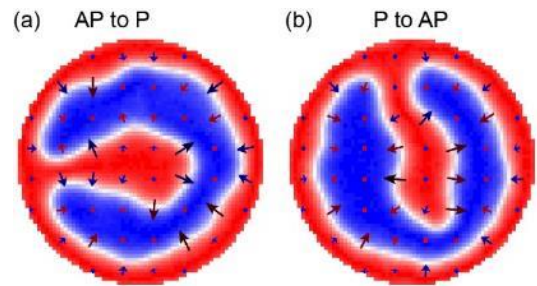

However, the disorder of magnetic textures resulting from the coexistence of DMI and STT becomes more obvious in a larger MTJ. Once the DMI is larger than the critical value, i.e, D>D$_C$, a curly domain occurs, as shown in Fig. S11.

*Figure S11 The magnetic state in the FL of an R=120nm MTJ after applying a current $|j_z| = 8 \times 10^{10}$ A/m$^2$ for 10 ns; D=1.75 J/m$^2$ for both cases. The corresponding stray field used is shown in Fig. S9.*

## 3. How to obtain a multi-skyrmions state in an MTJ ?

From the above simulations, a multi-skyrmions state can be obtained when both the applied current and the DMI are very large. Here, we simulated the reversal of a R=120 nm MTJ with D =

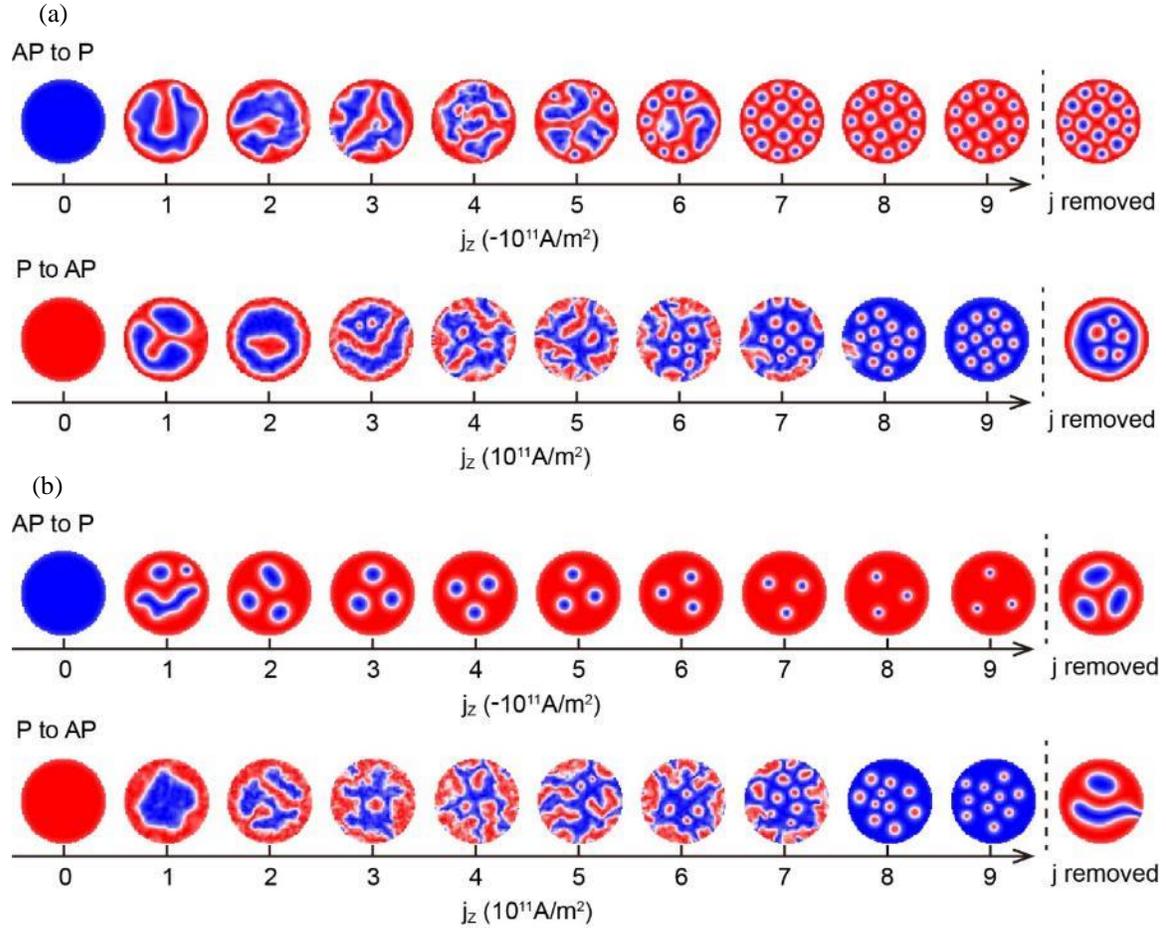

*Figure S12 The magnetic reversal the FL a MTJ with R=120 nm. D=1.75 J/m². The current changed step by step. For each step, the current was applied continuously for 10ns and each image was snapshot at the end of this duration. After obtaining stable skyrmions with high current at the end of the simuation, the current was removed. The last images show the stable state after removing the current. (a) Stray fields used corresponds to that in Fig.S9 and FL-STT was not taken into account. (b) Stray fields used corresponds to that in Fig.S13 and FL-STT was taken into account.*

1.75 mJ/$m^2$. The stray field considered is shown in Fig. S9. For both reversal directions, as the applied current increases, a disordered spin texture appears. When the applied current further increases, the stable skyrmions lattice appears. The stabilization of this skyrmion lattice is similar to the field stabilized skyrmion lattice. In a material with D>$D_C$, a mazy domain texture is stable in the ground sate. Many experiments demonstrated that a strong external field could change this mazy domain state to the skyrmions lattice state. Note that, here, in the reversal process shown in Fig. S12 (a), no FL-STT was considered. It demonstrates that a DL-STT can stabilize the skyrmion lattice. Moreover, the strongly disordered spin texture caused by the STT ($|j_z| = 4 \sim 6 \times 10^{10}$ A/$m^2$) helps to form a more homogenous skyrmions lattice ($|j_z| \geq 7 \times 10^{10}$ A/$m^2$).

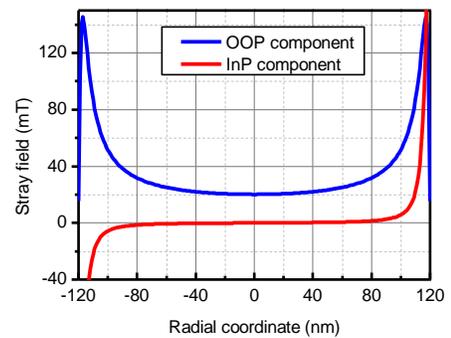

*Figure S13 The distribution of the stray field in the FL of a R=120nm MTJ. Structure of PL: SPL(1.5nm)/RL(3nm)/HL(1nm).*

Compared to the P to AP reversal, we can find that the current required for the stabilization of skyrmion lattice is a little smaller for the AP to P reversal. This may be explained by the fact that the strong stray field in the edge of FL points to +z direction, helping to terminate the disordered spin texture state and stabilize the skyrmions lattice. One can find that the skyrmions close to the edge were first stabilized. On the contrary, in the P to AP reversal, because of the competition between STT and $B_S$ in the edge, the magnetic state is more disordered before forming the skyrmions lattice. At last, the skyrmions in the center were first stabilized.

One may curious that whether the current required for the stabilization of the skyrmion lattice in an MTJ can be reduced. Now, we modify the stray field, as shown in Fig. S13. A biased stray field can be obtained by reducing the thickness of the HL. Then, the reversal of FL was simulated, as shown in Fig. S12(b). Note that, exceptionally, we considered the FL-STT in this simulation ($\tau_{FL} \sim 20\% \cdot \tau_{DL}$). We can find that the skyrmions can be stabilized with a smaller current in the AP to P reversal. However, less skyrmions were created compared with the former case. For the P to AP reversal, a large current is still required to stabilize the skyrmions lattice.

## V. About the magnetization - current hysteresis loop

In the main text, we gave a magnetization - current hysteresis loop, in which each $\bar{M}_Z$ was acquired after a relaxation of 10 ns (i.e, the current was stopped for 10 ns between tow current steps). Here, we give another magnetization - current hysteresis loop. The current was applied continuously in a stepwise manner. The $\bar{M}_Z$ was acquired at the end of each current steps, as shown in Fig. S14.

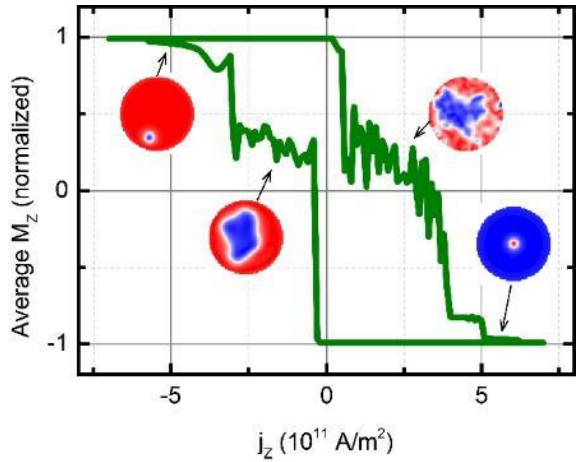

*Figure S14 A magnetization - current hysteresis loop of a R=60nm MTJ obtained based on the micromagnetic simulation. Current changed step by step and each step lasts for 10 ns. $D = 1.0 \, mJ/m^2$ and the stray field used corresponds to that in the moderate one in Fig. S2.*